\documentclass[12pt,a4paper,final]{iopart}

\usepackage{xcolor}

\usepackage{iopams}  
\usepackage{graphicx}
\usepackage[breaklinks=true,colorlinks=true,linkcolor=blue,urlcolor=blue,citecolor=blue]{hyperref}

\begin{document}

\title[Topological spin waves in the atomic-scale magnetic skyrmion crystal]{Topological spin waves in the atomic-scale magnetic skyrmion crystal}

\author{A. Rold\'an-Molina}
\address{Centro para el Desarrollo de la Nanociencia y la Nanotecnolog\'ia, CEDENNA, Avda. Ecuador 3493, Santiago 9170124, Chile}
\address{Departamento de F\'isica, Facultad de Ciencias F\'isicas y Matem\'aticas, Universidad de Chile, Casilla 487-3, Santiago, Chile}
\address{Instituto de F\'isica, Pontificia Universidad Cat\'olica de Valpara\'iso, Avenida Universidad 330, Curauma, Valpara\'iso, Chile}
\ead{alroldan.m@gmail.com}

\author{A.  S. Nunez}
\address{Departamento de F\'isica, Facultad de Ciencias F\'isicas y Matem\'aticas, Universidad de Chile, Casilla 487-3, Santiago, Chile}
\ead{alnunez@dfi.uchile.cl}

\author[cor1]{J.  Fern\'andez-Rossier\footnote{Permanent Address: Departamento de F\'isica Aplicada, Universidad de Alicante}}
\address{International Iberian Nanotechnology Laboratory, Av. Mestre Jose Veiga, 4715-310 Braga, Portugal}
\eads{\mailto{joaquin.fernandez-rossier@inl.int}}

\begin{abstract}
We study the spin waves of  the triangular skyrmion crystal  that emerges in a two dimensional spin lattice model as a result of the competition between Heisenberg exchange,  Dzyalonshinkii-Moriya interactions,  Zeeman coupling and uniaxial anisotropy.  
The calculated  spin wave bands have a finite Berry curvature that,  in some cases, leads to non-zero   Chern numbers, making this system  topologically
distinct from conventional magnonic systems. 
We compute  the   edge spin-waves, expected from the bulk-boundary correspondence principle, and show that they are chiral, which makes them immune to elastic backscattering. Our results illustrate how  topological phases can occur in self-generated emergent superlattices  at the mesoscale. 

\end{abstract}

\pacs{00.00, 20.00, 42.10}
\vspace{2pc}
\noindent{\it Keywords}: topological magnonics, skyrmion lattice
\submitto{\NJP}

\section{Introduction}

One of the major breakthroughs in the last few decades in condensed matter physics is the notion of  topological order
 associated to electronic states in crystals, including quantum Hall\cite{VK}, quantum spin Hall insulators\cite{Kane-Mele1,Kane-Mele2} and quantum anomalous Hall insulators\cite{Haldane88}  in two dimensions, and topological insulators in three dimensions\cite{TI3D}.  A specially important  aspect of all these topological phases in crystals is the existence of in-gap edge states that are immune to backscattering. This turns out in the control of anomalous transport properties, among which  the quantized Hall effect\cite{VK} is the most outstanding. In all these cases it is possible to define a topological invariant, associated to a mapping of the reciprocal space inherent to the crystal and Bloch states. 

Whereas these topological phases were initially proposed for electrons,  it became evident that other waves, such as photons\cite{Haldane08,Wang08,Wang09},  polaritons\cite{Karzig15}  and spin waves\cite{Murakami2013,Li2013,Mook2014,Mook2015} can also be  topological, and there have been both theoretical proposals, as well as experimental demonstration in some instances\cite{Wang09}.   In most of the previous works  the crystals that host  these topological phases belong to two types:  naturally occurring  solids\cite{Chisnell15} and patterned artificial structures or metamaterials\cite{Wang08,Murakami2013}. 
For instance, topological spin waves have been predicted to occur in mesoscopic crystals with artificial patterned ferromagnetic structures \cite{Murakami2013}, and also in atomic scale magnetic insulators as the Kagome lattice \cite{Li2013,Mook2014,Mook2015}, for which recent neutron scattering experiments give indirect evidence\cite{Chisnell15}.  
 Here we show that  topological spin waves can also be found    at an intermediate scale, in self-generated 
  skyrmion lattices that emerge as a non-trivial super-structure in the atomic  crystal.  (see figure 1).

Magnetic  skyrmions are non-coplanar  spin textures characterized by a non-zero  winding number $N$ associated to the mapping  defined by the magnetization field $\vec{M}(\vec{r})$ \footnote{
In 2D the skyrmion winding number is defined as 
$N= \frac{1}{4\pi}\int\vec{n}\cdot\left(\frac{\partial \vec{n}}{\partial x}
\times\frac{\partial \vec{n}}{\partial y}\right) d A$
where $\vec{n}=\frac{\vec{M}}{|\vec{M}|}$ is the unit 
vector associated to the skyrmion magnetization. 
}. 
%
%
In the last few years there has been a revival  in the interest of magnetic skyrmions \cite{Rossler06,Muhlbauer,Neubauer,Jonietz,Yu10,Heinze,Seki,Fert,Romming2013,Nagaosa-Tokura2013,Romming2015}  fuelled by  the observation of skyrmion crystals both in non-centrosymmetric compounds such as  MnSi\cite{Muhlbauer} and Cu$_2$OSeO$_3$ \cite{Seki} and in artificial atomically thin multilayers\cite{Heinze,Romming2013} and the promise of possible applications in spintronics due to the low current necessary to move them\cite{Fert,Nagaosa-Tokura2013}.  

As in every other broken symmetry  magnetic ground state\cite{Auerbach,Yosida},  spin waves excitations are expected to play an important role in the small energy spin dynamics of the skyrmion lattice\cite{Batista14,Roldan15}.  More generically, the question of how the non-collinear magnetic ground states influences the dynamics of the spin waves has been explored by several authors in the past \cite{Sheka04,Bruno,Ivanov07,Hoogdalem,Elias,Oh}. Thus,  Dugaev {\em et al.}\cite{Bruno}, realized that spin waves travelling in a background of non-homogeneous magnetization could acquire a Berry phase that in some instances would  affect their motion exactly a  magnetic field  affects charged particles.
This notion has been further explored and  confirmed in the context of spin waves of skyrmions and other topological defects \cite{Ivanov07,Elias,Oh,Hoogdalem}. It connects very well with earlier work  that found out that electrons surfing non-coplanar magnetic configurations would also be affected by an effective Lorentz force\cite{Loss90} resulting in a contribution to the anomalous Hall response\cite{Taguchi2001,Tatara02}, that could even result in quantum anomalous Hall phases\cite{Ohgushi00,Martin08}.  Recent work has shown that 
 spin waves
\cite{Nagaosa2014,Garst14} are known to undergo skew scattering by a  single skyrmion,  which implies the existence of an effective Lorenz force. In addition, skyrmion lattices are known to induce the quantum anomalous Hall phase in electrons\cite{NagaosaQH15,Lado2015},    motivating  our exploration of  the topological properties of the spin waves in a skyrmion lattice. 

\section{Model Hamiltonian}
  We consider   a two dimensional triangular lattice of spins,  inspired by the observation of a skyrmion lattice in a monolayer of Fe(111) on top of Iridium\cite{Heinze}. The Hamiltonian has   a single ion uniaxial anisotropy term that favors  the magnetization along the $z$ axis, 
 a first neighbor ferromagnetic exchange $J$,  the  Dzyaloshinskii-Moriya  interaction $D$ and the Zeeman term\cite{Batista14,Roldan15,Iwasaki2,ARM1}:
  \begin{eqnarray}
H&=&-\sum_{<i,j>}J_{i,j}\vec{S}_{i}\cdot \vec{S}_{j}+\sum_{<i,j>}\vec{D}_{i,j}\cdot(\vec{S}_{i}\times \vec{S}_{j})- \sum_{i}\vec{h}\cdot\vec{S}_{i}-K\sum_{i}(S_{i}^{z})^{2},\nonumber \\
\label{Hamiltonian1}
\end{eqnarray}
In a first step,  we approximate  the spins as classical objects.  We define a classical functional 
${\cal E}_{cl}\equiv H(\vec{\Omega}_i)$,
where the spin operators $\vec{S}$ are replaced by classical vectors $\vec{\Omega}_i$. 
The classical ground state is  defined as the configuration $\vec{\Omega}_i$ that minimizes the functional ${\cal E}_{cl}$. We find it by  self-consistent iteration.

\begin{figure}
\begin{center}
\includegraphics[width=0.8\textwidth]{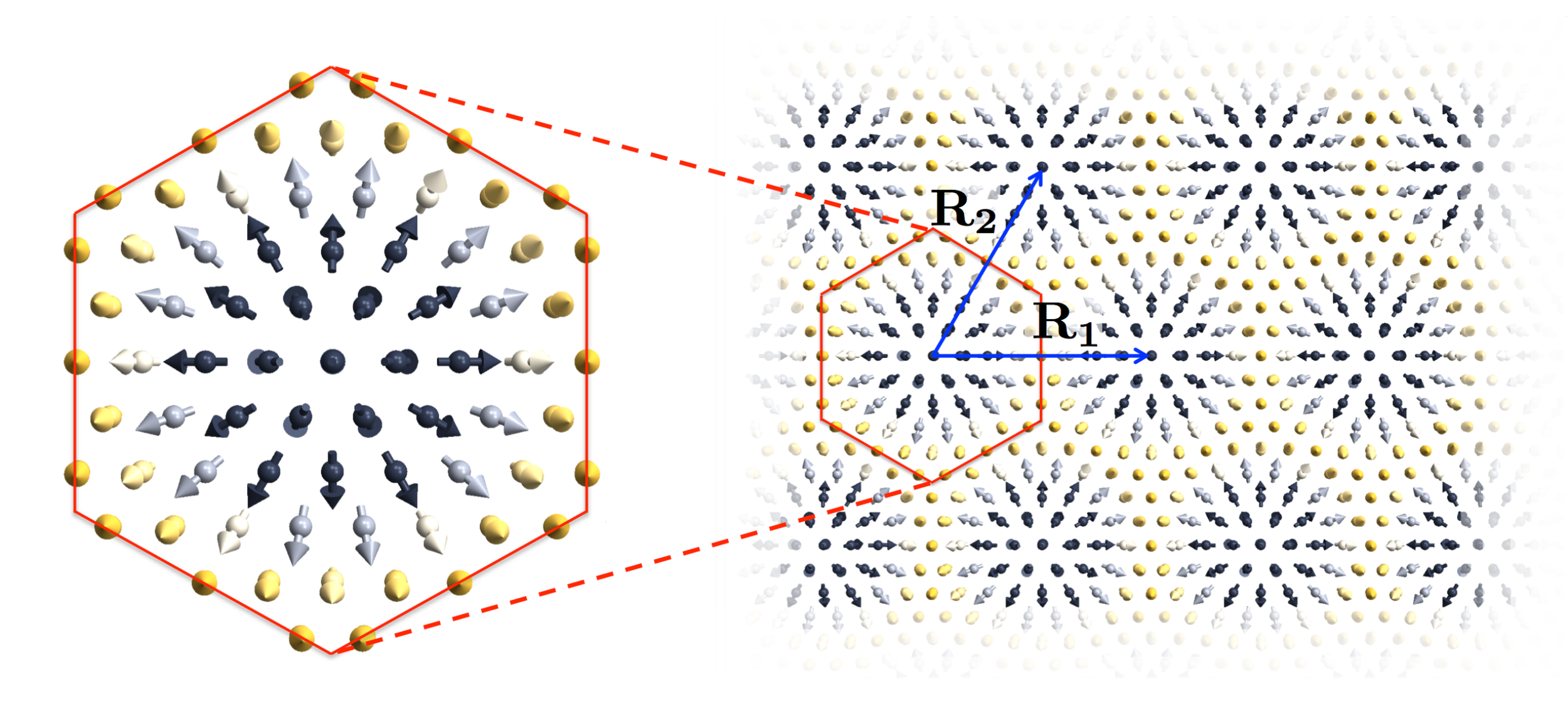}
\end{center}
\caption{Triangular skyrmion lattice for a classical spin system described by the Hamiltonian (\ref{Hamiltonian1}) with $J=1[meV]$, $D_0=1[meV]$, $K=0.5[meV]$ and $h=0.36(D_{0}^2/J)$. The form of the Dzyalonshinskii-moriya interaction that we have chosen favors N\'eel-like skyrmions. As it can be seen in the inset within each unit cell rim spins (yellow)  are antiparallel to core spins (black). }
\label{fig: sklattice}
\end{figure}

The non-coplanar  spin alignment between  first neighbors is the result of the competition between the standard
Heisenberg interaction ($-J\vec{S}_{i}\cdot \vec{S}_{j}$) and the antisymmetric Dzyaloshinskii-Moriya interaction (DMI) term $\vec{D}_{i,j}\cdot\vec{S}_{i}\times \vec{S}_{j}$, that requires breaking of inversion symmetry.  
Here we take a DMI compatible with interfacial inversion symmetry breaking,  $\vec{D}=D_0 (\hat{z}\times\vec{n}_{ij})$, 
where $\vec{n}_{ij}$ is the unit vector that joins sites $i$ and $j$, 
$J=1[meV]$, $D_0=1[meV]$, $K=0.5[meV]$ .
At zero $h$, the ground state is an helimagnet.
  Application of a magnetic field $\mathbf{h}$ along the $\hat{z}$ direction,  perpendicular to the system surface and thus to all the $\vec{D}_{ij}$ vectors, 
  favors the formation of non-coplanar Skyrmions lattice, shown in figure (\ref{fig: sklattice}) for $h=0.36(D_{0}^2/J)$.  It is apparent that the classical ground state is a triangular lattice of  skyrmions, i.e., non coplanar structures with a core spin pointing in a direction opposite to the interstitial spins.

\section{Topological magnonic bands}
\begin{figure}
\begin{center}
\includegraphics[width=0.6\textwidth]{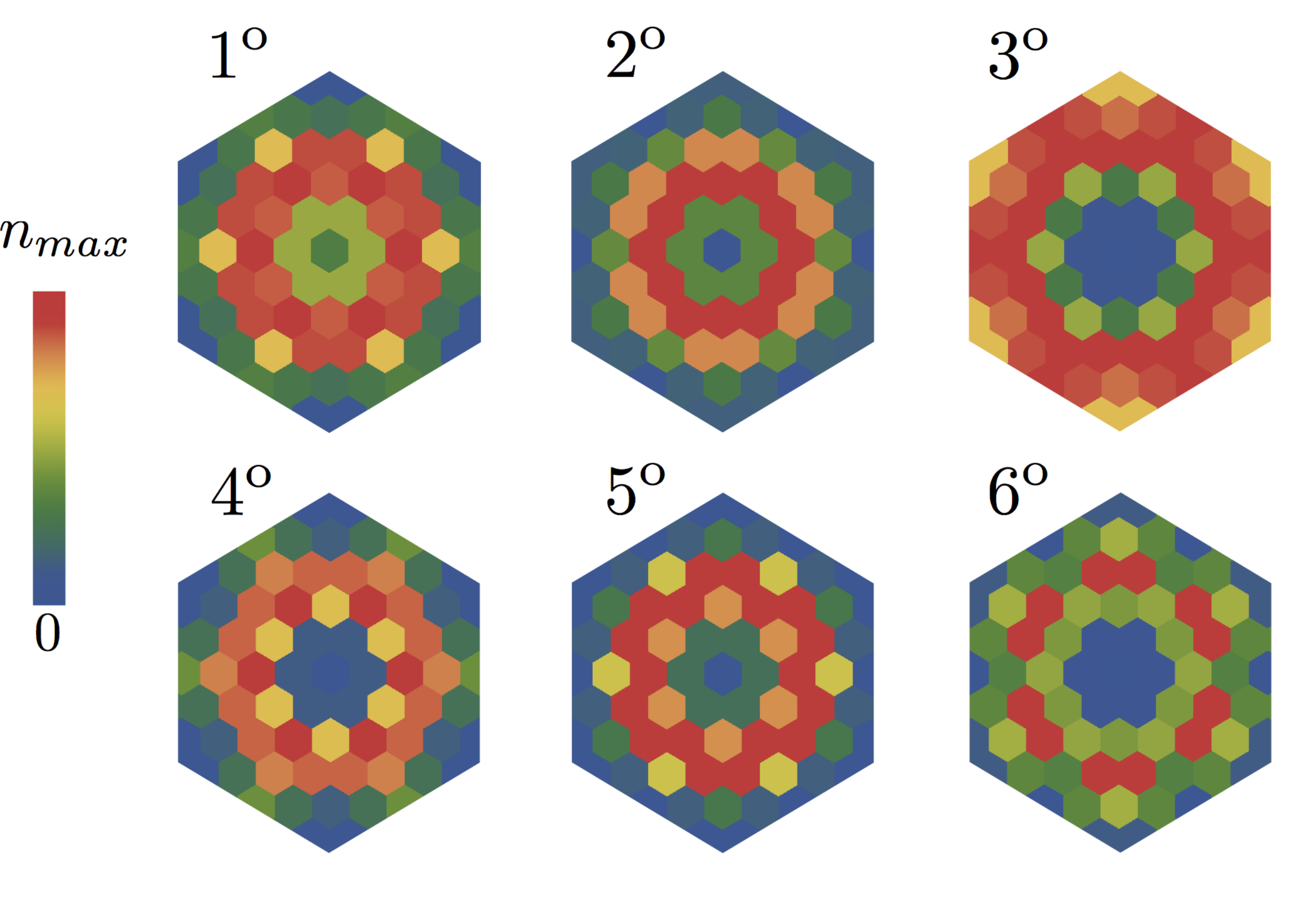} 
\end{center}
\caption{ Magnonic occupations within the unit cell evaluated at $\Gamma$ point for the six lowest bands. A clear distinction can be made between the modes bounded to the skyrmion  and the extended modes. The different modes can be related to the ones in the case of an isolated skyrmion. }
\label{fig: Magnon Occupation}
\end{figure}

\begin{figure}[hbt]
\begin{center}
\includegraphics[width=0.7\textwidth]{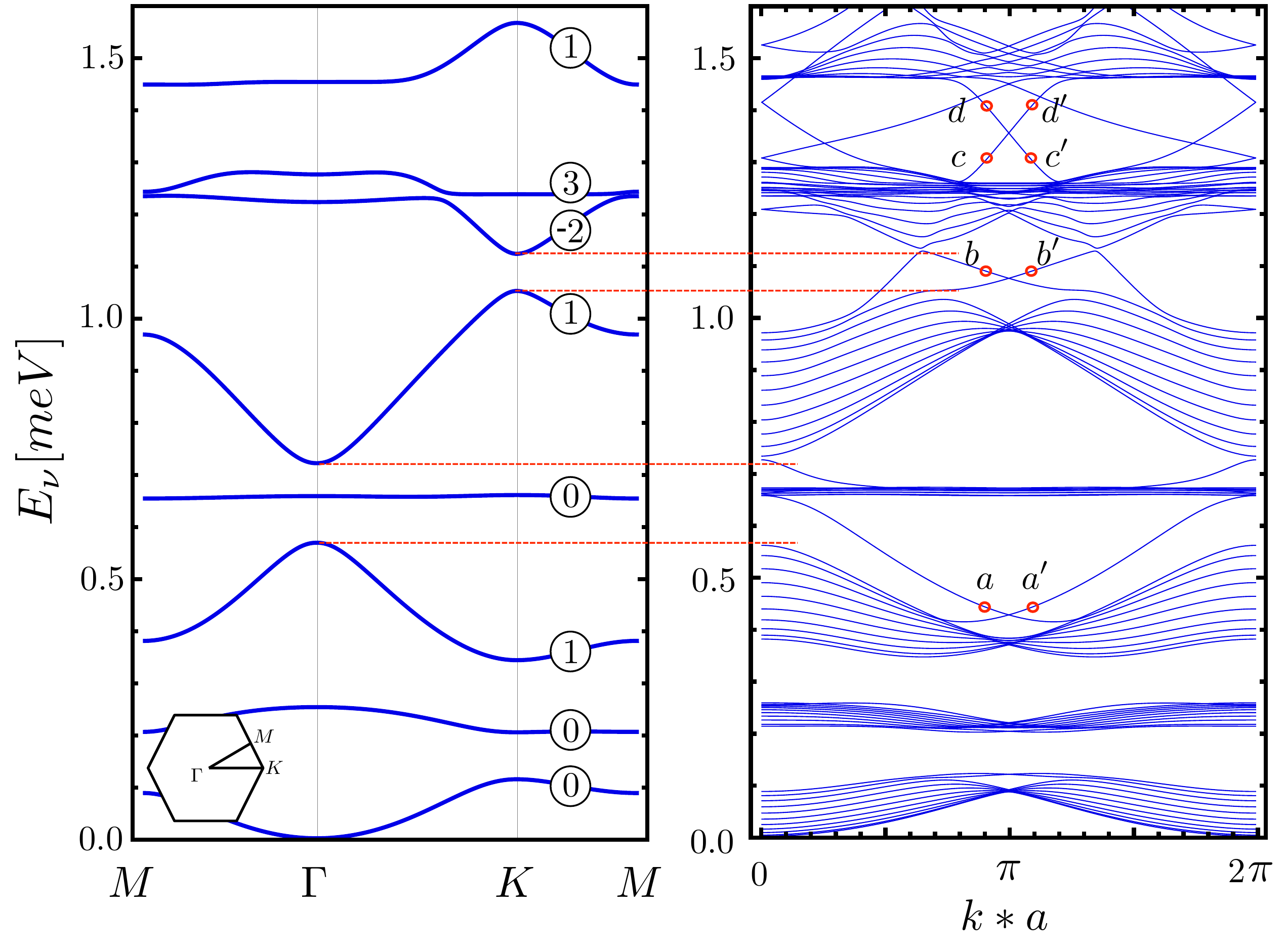}
\caption{ Left panel: Spin wave dispersion for two dimensional crystal shown in figure 1. The momenta are selected along the path connecting the points of high symmetry (see inset). For each band we have calculated a corresponding Chern number, Eq.(\ref{eqn: Chern Number}),  that is shown for each band.
Right panel: spin wave bands for a one dimensional  strip 11 skyrmions wide. Each one of the bands of the homogeneous crystal is split into 11 sub-bands. The modes crossing over the gaps are unidirectional and topologically protected. }
\label{fig: spectra}
\end{center}
\end{figure}

The skyrmion crystal also has spin waves,   like in the case of  other symmetry breaking magnetic states, such as ferromagnetic and antiferromagnetic states \cite{Batista14,Roldan15}.
Here we describe them using a conventional quantum mechanical   Holstein Primakoff (HP) boson theory\cite{Auerbach,Yosida,Oh, Roldan15, ARM1,HP} for spin waves: 
\begin{eqnarray}
\mathbf{S}_i\cdot\vec{\Omega}_i&=&S-n_i,\,\, \nonumber \\
  S_{i}^{+}&=&\sqrt{2S-n_{i}}\ a_{i}, 
 \,\,S_{i}^{-}=a^{\dagger}_{i}\sqrt{2S-n_{i}}\ ,
\end{eqnarray}
where $\vec{\Omega}_{i}$ is the spin direction of the classical ground state, obtained above,  on the position $i$, $a_i^{\dagger}$ is a Bosonic creation operator and $n_i=a^{\dagger}_ia_i$ is the boson number operator.  
The spin-wave approximation consists on keeping only quadratic terms in the HP bosons, when  Hamiltonian  (\ref{Hamiltonian1})
is written in terms of  HP bosonic  operators\cite{Auerbach,Yosida}. 
In this context  we approximate $S_{i}^{-}=\sqrt{2S}\ a_{i}^{\dagger}$. Thus,  the annihilation of a  HP  boson is equivalent the addition of one unit of spin angular momentum from the classical ground state. Since we are dealing with a crystal, it is convenient to work in the reciprocal space.  We label the sites in the crystal by a  unit cell vector  $\vec{R}$ and, inside each unit cell, by an additional label $j$.  We define
\begin{center}
\begin{eqnarray}
a_{\vec{R},j}=\sum_{\vec{k}}e^{i \vec{R}\cdot\vec{k}} \ {a_{\vec{k},j}}
\end{eqnarray}
\end{center}
After some algebra, the spin wave Hamiltonian for the skyrmion crystal reads:
\begin{eqnarray}
H(\bf{k})=\left[\bf{a}_{\bf{k}}^{\dagger}\ a_{-\bf{k}}\right] 
\left[
\begin{array}{cc}
\bf{H}(\bf{k}) & \bf{\Delta}({\bf{k}})\\
\bf{\Delta}^*(-\bf{k}) & \bf{H}^*(-\bf{k})
\end{array}
\right]
\left[
\begin{array}{cc}
\bf{a}_{\bf{k}} \\ \bf{a}^{\dagger}_{-\bf{k}}
\end{array}
\right],
\label{Hamiltonian2}
\end{eqnarray}  
where $\bf{a}_{\bf{k}}^{\dagger}$ is a vector of HP bosons of dimension $N$, given by the number of spins in the unit cell of the crystal and $\bf{H}$ and $\bf{\Delta}$ are matrices, functional of the classical ground state\cite{ARM1,Roldan15}.  As in other cases with non-collinear ground states,   the HP Hamiltonian contains terms that do not commute with the boson number operator.  This Hamiltonian can be diagonalized, using a paraunitary transformation\cite{Colpa,ARM1,Roldan15}, leading to:
\begin{equation}
H(\vec{k})= \sum_{\nu,\vec{k}} E_{\nu}(\vec{k})  \left(\alpha^\dagger_{\nu,\vec{k}}	\alpha_{\nu,\vec{k}}+\frac{1}{2}
\right)
\end{equation}
$\nu=1,N$  labels the spin wave bands, and $\vec{k}$ lives in the first Brillouin zone.  Our results for the two dimensional skyrmion lattice are shown in Fig. (\ref{fig: spectra}) for the lowest energy bands.  It is apparent that we obtain a set of non-overlapping spin wave bands separated by gaps.  Similar results have been obtained for a skyrmion square lattice using the same method\cite{Roldan15} and using a long-wave length approximation for the triangular skyrmion lattice\cite{Hoogdalem}. The non-overlapping bands  can be related to the various lowest energy spin wave modes associated to the spin wave excitations of a single skyrmion\cite{Roldan15}.   We can map their spatial distribution over the unit cell as the average of the HP boson occupation, which measures the departure of the magnetization from the classical ground state :
\begin{equation}
\langle n_j\rangle \equiv \langle \psi_{\nu}(\vec{k}) |a^{\dagger}_j a _j|\psi_{\nu}(\vec{k})\rangle
\end{equation}
where $|\psi_{\nu} \rangle= \alpha^{\dagger}_{\nu,\vec{k}}|G\rangle $ are the spin wave functions and $|G\rangle$ is
the ground state of eq. (\ref{Hamiltonian2}).      Results for the spin wave modes are shown in Fig. (\ref{fig: Magnon Occupation}).

\begin{figure}
\begin{center}
\includegraphics[width=0.6\textwidth]{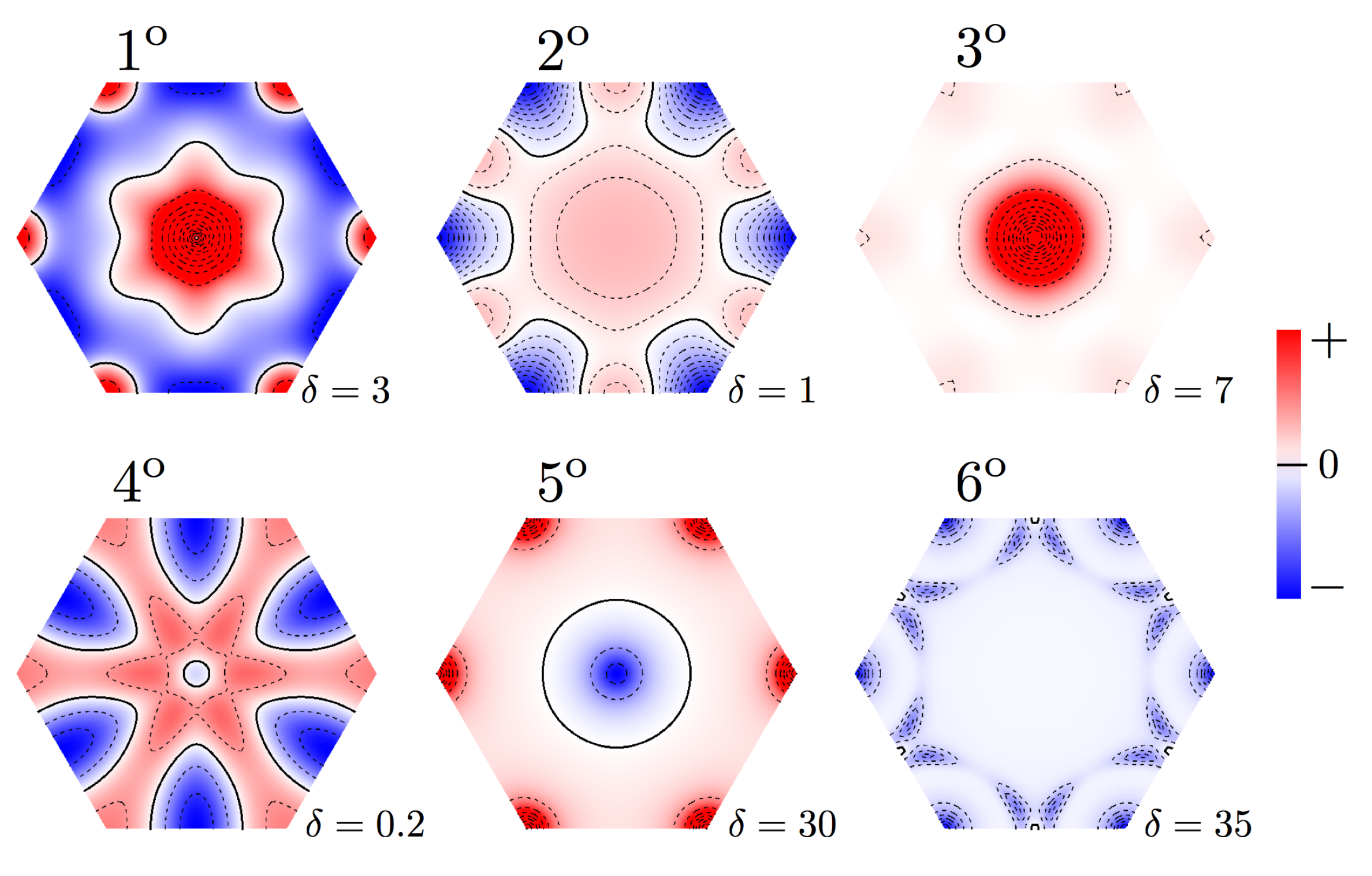}
\end{center}
\caption{ Berry curvature for the six lowest energy bands on the Brillouin zone. The curvature is calculated as indicated in Eq. (\ref{eq: Berry Curvature}). Level curves have been chosen taking different height intervals $\delta$ indicated in  each case. For reference the black line denotes zero curvature. }
\label{fig: Berry Curvature}
\end{figure}

We can assign a Berry curvature to every spin-wave band \cite{Murakami2013}:
\begin{eqnarray}
\Omega_{\mu}(\vec{k})=
 \sum_{\nu\neq \mu}\frac{(-1)^{\sigma^\nu} (-1)^{\sigma^{\mu}}}{(E_{\nu,\bf{k}}-E_{\mu,\bf{k}})^2} \left( \langle \psi_{\mu}| \frac{\partial H}{\partial k_{x}} | \psi_{\nu} \rangle \langle \psi_{\nu}| \frac{\partial H}{\partial k_{y}} | \psi_{\mu} \rangle -(x \rightleftarrows y) \right) 
\label{eq: Berry Curvature}
\end{eqnarray}
where $\sigma_\mu$ is 0 (1) for positive (negative) energy bands \cite{Murakami2013}. 
In Fig. (\ref{fig: Berry Curvature}) we display the Berry curvature for the six lowest bands evaluated at each point of the Brillioun zone.
 \newpage
The Chern number, the integral of the Berry curvature,
\begin{eqnarray}
{\cal C}_{\mu}=\frac{i}{2\pi}\int_{BZ}d^2{\bf k}\; \Omega_{\mu}(\vec{k})
\label{eqn: Chern Number}
\end{eqnarray}
is non zero for some of the bands ${\cal C}_{\mu}\neq 0$.  This is the central result of this work.  More specifically, for the spin-wave spectrum in figure (\ref{fig: spectra}), the sequence of Chern numbers, in ascending order,  is $0,0,+1,0,+1,-2,3,1,..$.  We have verified that the sum of Chern numbers over the entire spectrum is zero, as expected\cite{Mook2015}.
  We have  also found that the sign of the Chern number is controlled by the sign of the applied magnetic field. Changing the sign of $D$ results in a change of the skyrmion winding numbers, but it does not change the sign of the  spin-wave Chern numbers, in contrast with the case of the electronic Chern number in a skyrmion lattice\cite{Lado2015}.  
 \begin{figure}[hbt]
\begin{center}
\includegraphics[width=0.7\textwidth]{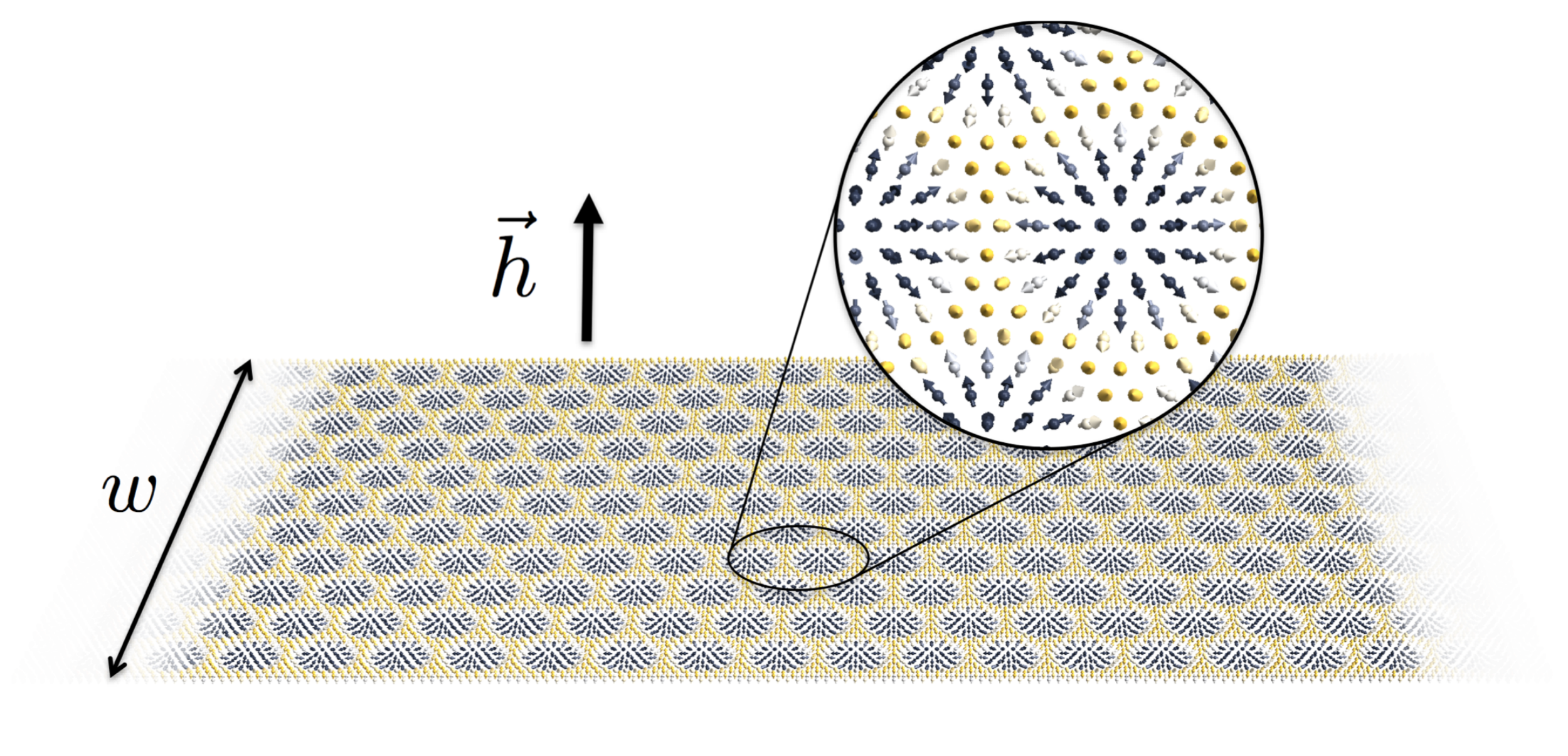} 
\includegraphics[width=0.6\textwidth]{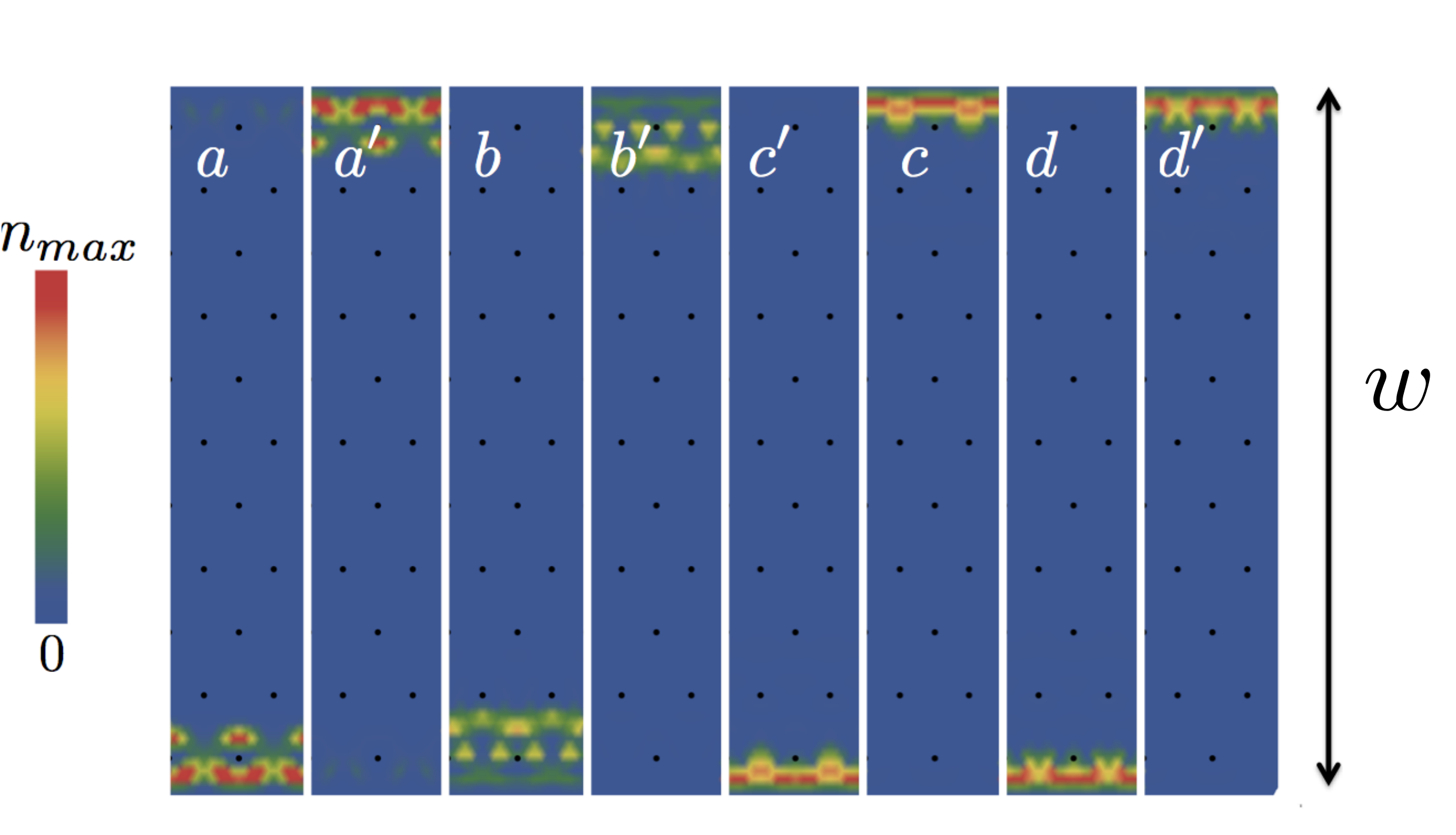}
\end{center}
\caption{Top: Strip geometry of 11 skyrmions wide. We have calculated the dispersion relation for this system. The results are shown, as function of the longitudinal momenta, in Fig. (\ref{fig: spectra}).  Bottom: Magnon density for the edge modes $a$, $a^\prime$, $b$, $b^\prime$, $c$, $c^\prime$, $d$, and $d^\prime$ modes depicted in Fig. (\ref{fig: spectra}). The black dots correspond to the centers of the underlying skyrmions. It is evident that those chiral modes correspond to edge excitations. }
\label{fig:edge states}
\end{figure}

\section{Magnonic edge states}
 The fact that ${\cal C}_{\mu}\neq 0$ automatically entails non-trivial consequences for the edge states of the system that we explore in the rest of this work.   For that matter, it is convenient to define  the so-called\cite{Mook2014} winding number of a given band-gap $\mu$  as the sum of all the Chern numbers of the bands up to band $\mu$ :
 \begin{equation}
 \nu_{\mu}\equiv\sum_{\mu'=1,\mu}{\cal C}_{\mu'}
 \end{equation}
According to the bulk-edge correspondence principle\cite{Hatsugai93},   the number of in-gap   one-way edge states,  is determined by $\nu_{\mu}$.  More specifically, 
the interface between two insulators $A$ and $B$,  with  winding number  numbers $\nu_A$ and $\nu_B$ hosts $|\nu_A-\nu_B|$ interface states.    

In order to study the properties of these in-gap  states we consider a one dimensional strip with a finite width of 11 unit cells (i.e., 11 skyrmions) and up to  $97\times 8$ spins.
   The calculation of the spin waves follows the same procedure than in the infinite crystal. Rather than being inferred from the 2D solution the classical ground state of the strip is obtained numerically. Away from the edges, the classical solution is identical to the one in 2D, as expected (see Fig. \ref{fig:edge states}).   The resulting one dimensional bands are shown in the figure.   It is apparent that the two lowest energy  2D bands, with ${\cal C}=0$, give rise to  a group of 11 bands in the strip.

  When the winding number below a given gap is $\nu$, there are $|\nu|$  {\em in-gap} edge states
  (see for instance bands with labels $a,a',b,b',c,c',d,d'$).   Inspection of the average occupation of the HP boson associated to these states  
  reveals  that these in-gap states  are edge states indeed. Importantly,  it is apparent that states $a,b,c,d$, all of them chosen with $\frac{\partial E}{\partial k}$  opposite to those of $a',b',c',d'$ are localized in opposite edges.  In addition,  states with negative (positive)  velocity, such as  $a,b,c',d$  ($a',b',c,d'$) are all localized in the bottom (top) edge.   Thus,  the edge states can only propagate in one direction, in stark contrast with normal confined modes. Since these edge states occupy spectral regions in which the bulk spectrum is gapped, elastic backscattering is impossible for them in sufficiently wide stripes.

\begin{figure}[hbt]
\begin{center}
\includegraphics[width=0.75\textwidth]{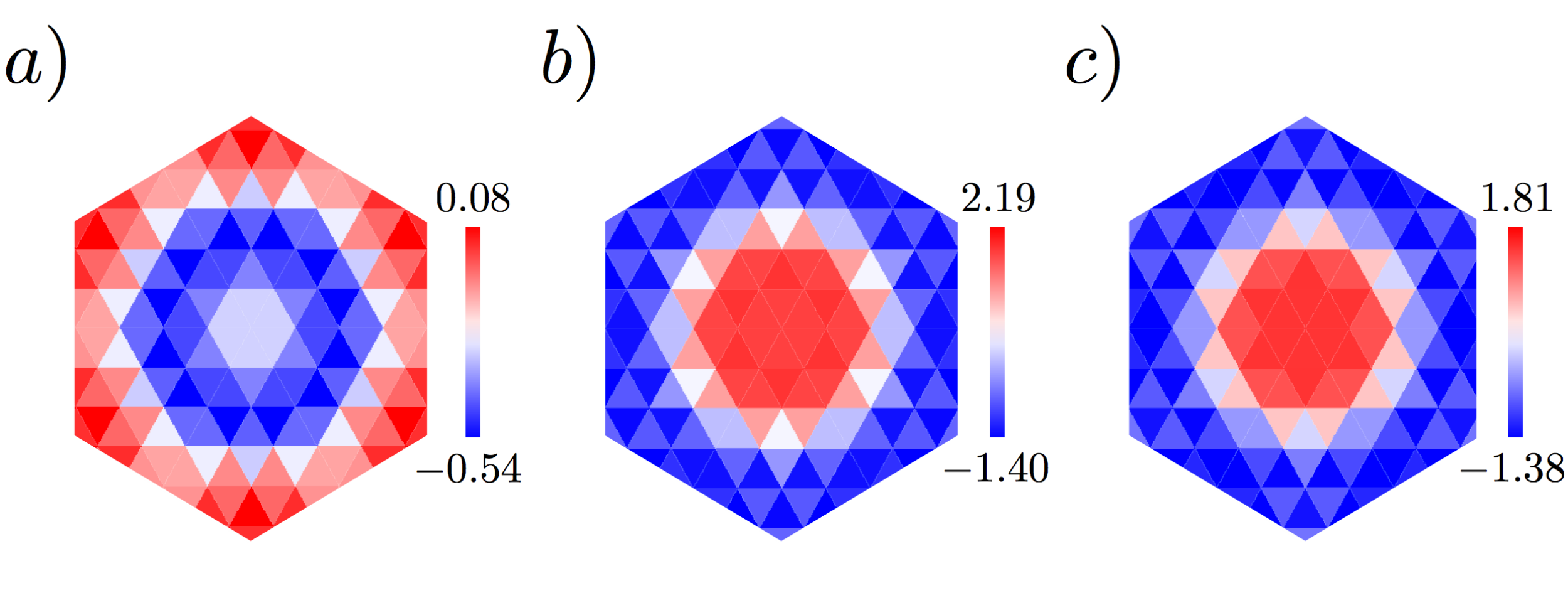}
\end{center}
\caption{Magnetic flux associated with both contributions (chirality, (a) vorticity (b) and total (c)). The total contribution off the vorticity (b) vanishes, so that  the net flux is a consequence of the chirality contribution only }
\label{fig: Magnetic Flux}
\end{figure}

\section{Origin of the anomalous spin wave force }
We now describe  quantitatively  the physical origin of the finite Chern number of the spin waves in the skyrmion crystal.   In analogy with the 
case of electronic quantum anomalous Hall insulators,  there has to be a Lorentz-type or anomalous force acting on the spin waves in the skyrmion crystal.  It must be noted that, although the Hamiltonian describing the skyrmion crystal is the similar to the one describing the Hall effect of spin waves in Lu$_2$V$_2$O$_7$\cite{Onose2010}, the origin of the anomalous force and the resulting spin-wave Hall effect in these systems is actually quite different.  In the case of the Kagome ferromagnets,   the classical ground state is collinear so that the resulting   spin wave dynamics,  as described with the HP theory\cite{Li2013,Mook2014,Mook2015},  is mathematically identical to the one of electrons in the lattice.  
In the case of the spin waves in the Kagome ferromagnets, the role of the DM coupling is to provide a phase to the hopping terms\cite{Li2013,Mook2014,Mook2015}, very much like magnetic fields do on electrons in a lattice,  providing thereby the system with a finite flux across each triangular plaquette. 

In contrast,  the physics of the skyrmion-based topological magnonic crystal emerges from the Berry curvature brought about by the texture in the magnetization degree of freedom  \cite{Bruno,Elias,Oh,Hoogdalem}. 
This fact is better appreciated doing two approximations on our Hamiltonian. First, we take the continuum limit, regarding all variables as smooth functions of the space, slowly varying within the lattice constant scale. Additionally we take a classical limit, rendering all spin operators as simple vectors lying on the unit sphere. The resulting model is  similar to the system studied in \cite{Oh,Hoogdalem}.
In the continuum limit, Hamiltonian from Eq. (\ref{Hamiltonian1}) leads to the following 
 Hamiltonian density:
\begin{eqnarray}
H=\frac{3}{2}a^2 S^2 J\sum_{\mu}\partial_{\mu}{\bf{m}}\cdot\partial_{\mu}{\bf{m}}+3 a S^2 D\sum_{\mu}(m_{z}(\partial_{\mu}m_{\mu})-m_{\mu}\ (\partial_{\mu}m_{z}))\nonumber\\
-h S({\bf{m}}\cdot\hat{z})-KS^2({\bf{m}}\cdot\hat{z})^{2}
\end{eqnarray}
where $\mu=x,y$. Folllowing previous work \cite{Bruno, Elias, Oh, Hoogdalem}, it can be seen that the physics of small disturbances, $\delta \bf m$, about the equilibrium  magnetization texture, $\bf m_0$, isÊ dominated by an effective Hamiltonian with an O(2)-gauge symmetry. The vector potential associated with this effective Hamiltonian is\cite{Oh}:
\begin{eqnarray}
\bf{A}&=&(1-\cos{\theta})(\nabla\phi)+\kappa\ \hat{z}\times \bf{m}_0
\end{eqnarray}
%
 where $\kappa=2D/aJ$. In the latter equation $\theta$ and $\phi$ correspond to the polar angles of the equilibrium magnetization.
 The effective magnon hamiltonian correspond to an effective electron-like model where each hexagonal plaquette is pierced by a magnetic field equal to:
 \begin{eqnarray}
{\bf B}_{\rm eff}=(\nabla\times{\bf{A}})_{z}={\bf{m_{0}}}\cdot\left(\partial_{x}{\bf{m_{0}}}\times \partial_{y}{\bf{m_{0}}}\right)+\kappa\ (\partial_{x}m_{0}^{x}+\partial_{y}m_{0}^{y})
 \end{eqnarray}
 The first contribution to this effective magnetic field arises from the chirality of the underlying magnetic texture and is nothing but the density of topological charge whose net flux is equal to $4\pi$ within each unit cell. On the other hand, the contribution of the DM interaction is proportional to the divergence of the magnetization texture and its integral vanishes.  We have calculated both contributions separately within a unit cell of the system. The calculation is done for the equilibrium magnetization, Fig. (\ref{fig: sklattice}), and doing a barycentric interpolation of the discrete lattice. The results are shown in Fig. (\ref{fig: Magnetic Flux}).

\begin{figure}[hbt]
\begin{center}
\includegraphics[width=0.9\textwidth]{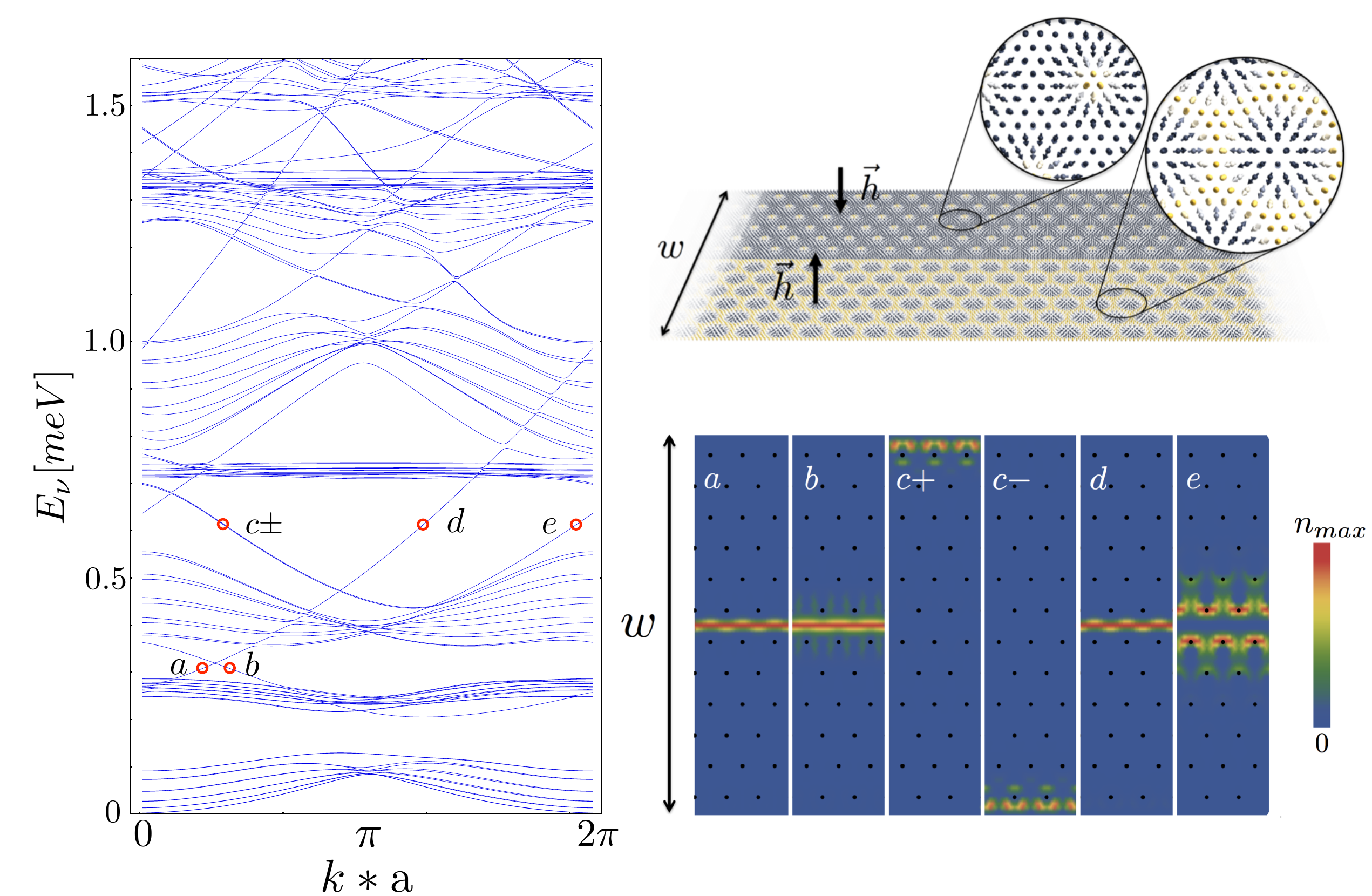}
\end{center}
\caption{Interface states created at a domain wall. Upper right: domain wall of opposite magnetic field leads to skyrmion crystals of opposite winding numbers. This is seen from the magnon system as an edge that connects regions with opposite Chern numbers. The domain wall gives rise to topologically protected interface modes. Left: Bands for the system. The topologically protected modes are highlighted with the labels  
$c$. States $a$ and $b$ are non-topological modes localized at the domain wall. $c^\pm$ highlights the edge states in the outer boundaries of the system. $d$ and $e$ are the unidirectional topologically protected modes living at the domain wall. Lower right: Magnon densities for the highlighted modes. The black dots correspond to the centers of the underlying skyrmions. }
\label{fig: interface}
\end{figure}

\section{Domain wall spin waves}

We conclude our work with a simple application of the previous ideas to create a tunable channel of unidirectional, topologically protected, spin waves. With this idea in mind we study consider the interface between two skyrmion crystals with opposite Chern numbers, in analogy with a similar calculation done by Mook {\em et al.} for the case of pyrochlore Kagome lattices\cite{Mook2015}.  In the case of skyrmions, the simplest way this can be achieved is through a domain wall that changes the sign of the magnetic field.  In our simulation, we consider a strip where the applied magnetic field is stepwise constant and has opposite sign in the top and bottom halves. The results are, nevertheless, robust to the analysis of more realistic domain walls. The resulting classical ground state, shown in Fig. (\ref{fig: interface}), has two skyrmion crystals with opposite winding number. The opposite winding number leads to opposite effective magnetic flux. Across  the interface region, where the magnetic field changes, there is a change in the sign of the Chern number that leads to topologically protected modes that propagate along the domain wall. 
Our calculations show that these states  connect otherwise separated bands (see  Fig. (\ref{fig: interface}).  This unidirectional  magnonic waveguides could be swhitched on and off by removing the domain wall, restoring for instance the homogeneous magnetic field.

\section{Conclusions}

 In summary, we have calculated the topological properties  for  spin waves in a  skyrmion crystal.   We have found that these bands have a finite Berry curvature, as expected from previous work\cite{Bruno, Hoogdalem} and we have computed their  Chern number, which   is non zero in several of them.    Thus,   a two dimensional skyrmion lattice would  realize the spin-wave analogue of the anomalous quantum Hall effect.  We have computed both the edge and interface states  of this system and verified that they comply with   the index theorem and,  in analogy with Landau-level edge states,  are unidirectional,  and thereby inmune to elastic backscattering.   To the best of our knowledge our results are the first example of a topological phase that occurs in a self-generated emergent meso-structure, such as a skyrmion lattice.   The peculiar properties of the edge topological edge states might find use in spintronics applications. 

\section{Acknowledgements}
The authors would like to thank funding from grants Fondecyt 1150072, ICM P10-061-F by Fondo de Innovaci\'on para la Competitividad-MINECON and Anillo ACT 1117. ASN also acknowledges support from Financiamiento Basal para Centros Cient\'ificos y Tecnol\'ogicos de Excelencia, under Project No. FB 0807(Chile).  ARM  and ASN  acknowledge hospitality of INL.   JFR acknowledges fruitful discussions with C. D. Batista and R. Wiesendanger during the SPICE workshop "Magnetic adatoms as building blocks for quantum magnetism".

\end{document}